\documentstyle[12pt,epsf]{article}
\begin{document}
\def\theequation{\arabic{section}.\arabic{equation}}
\def\Section#1{\setcounter{equation}{0}\section{#1}}

\newcommand{\ud}{\mbox{$U^{\dagger} $}}
\newcommand{\dm}{\mbox{$\nabla_{\mu}U $}}
\newcommand{\dmu}{\mbox{$\nabla^{\mu}U$}}
\newcommand{\dnu}{\mbox{$\nabla^{\nu} U$}}
\newcommand{\dn}{\mbox{$\nabla_{\nu}U $}}
\newcommand{\bu}{\mbox{$\bar U$}}
\newcommand{\bud}{\mbox{$\bar U^{\dagger} $}}
\newcommand{\dmd}{\mbox{$\nabla_{\mu}U^{\dagger }$}}
\newcommand{\dmud}{\mbox{$\nabla^{\mu}U^{\dagger } $}}
\newcommand{\bdm}{\mbox{$\nabla_{\mu} \bar U $}}
\newcommand{\bdmu}{\mbox{$\nabla^{\mu} \bar U $}}
\newcommand{\bdn}{\mbox{$\nabla_{\nu} \bar U $}}
\newcommand{\bdnu}{\mbox{$\nabla^{\nu} \bar U $}}
\newcommand{\bdmd}{\mbox{$\nabla_{\mu}\bar U^{\dagger} $}}
\newcommand{\bdmud}{\mbox{$\nabla^{\mu}\bar U^{\dagger } $}}
\newcommand{\bdnd}{\mbox{$\nabla_{\nu}\bar U^{\dagger} $}}
\newcommand{\bdnud}{\mbox{$\nabla^{\nu}\bar U^{\dagger } $}}
\newcommand{\dnd}{\mbox{$\nabla_{\nu}U^{\dagger} $}}
\newcommand{\dnud}{\mbox{$\nabla^{\nu}U^{\dagger } $}}
\newcommand{\qm}[1]{\mbox{$\nabla_{\mu}Q^{#1} $}}
\newcommand{\qmu}[1]{\mbox{$\nabla^{\mu}Q^{#1} $}}
\newcommand{\qn}[1]{\mbox{$\nabla^{#1}_{\nu}Q $}}
\newcommand{\qnu}[1]{\mbox{$\nabla^{#1\nu}Q $}}
\newcommand{\del}{\mbox{$\Delta_{\mu} $}}
\newcommand{\delu}{\mbox{$\Delta^{\mu} $}}
\newcommand{\bra}{\mbox{$\langle \, $}}
\newcommand{\ket}{\mbox{$\rangle  $}}
\newcommand{\field}{\mbox{$f_{\mu \nu} $}}
\newcommand{\fieldu}{\mbox{$f^{\mu  \nu} $}}
\newcommand{\gfield}[1]{\mbox{$F_{#1\mu \nu} $}}
\newcommand{\gfieldu}[1]{\mbox{$F^{\mu \nu}_{#1} $}}
\newcommand{\chid}{\mbox{$\chi^{\dagger} $}}
\newcommand{\paru}{\mbox{$\partial^{\mu} $}}
\newcommand{\parti}{\mbox{$\partial_{\mu} $}}
\newcommand{\fp}{\mbox{$F^{2}_{\pi} $}}
\newcommand{\fo}{\mbox{$F^{2}_{0} $}}
\newcommand{\bo}{\mbox{$B_{0} $}}
\newcommand{\ph}{\mbox{$a_{\mu} $}}
\newcommand{\phu}{\mbox{$a^{\mu}  $}}
\def\ar{\rightarrow}
\def\deh{{\pa}_{0}^{\!\!\!\!\leftrightarrow}\,}
\def\disc{\mbox{\,disc}\,}
\def\intf{\int d^{4}x\,}
\def\intgr{\int_{4}^{\infty}}
\def\ints{\int_{4}^{\infty} ds'}
\def\lar{\longrightarrow}
\def\pa{\partial}
\def\pal{p_{\al}}
\def\pbe{p_{\be}}
\def\pga{p_{\ga}}
\def\pde{p_{\de}}
\def\qba{\overline{q}}
\def\roo{(\frac{s-4}{4})}
\def\suz{\mbox{SU(2)}}
\def\sud{\mbox{SU(3)}}
\def\tp{\tilde p}
\def\tr{\mbox{tr}\,}
\def\Tr{\mbox{Tr}\,}
\def\ue{\mbox{U(1)}}
\def\uka{\underline{k}}
\def\upe{\underline{p}}
\def\al{\alpha}
\def\be{\beta}
\def\ga{\gamma}
\def\de{\delta}
\def\ka{\kappa\,}
\def\la{\lambda}
\def\ep{\varepsilon}
\def\om{\omega}
\def\Ph{{\it\Phi}}
\def\psiba{\overline{\psi}}
\def\si{\sigma}
\def\th{\theta}
\def\va{\varphi}

\def\beq{\begin{equation}}
\def\eeq{\end{equation}}
\def\bed{\begin{displaymath}}
\def\eed{\end{displaymath}}
\def\beqq{\begin{eqnarray}}
\def\eeqq{\end{eqnarray}}
\def\bedd{\begin{eqnarray*}}
\def\eedd{\end{eqnarray*}}
\def\Ph{{\it\Phi}}

\def\header{\begin{flushleft}
            ZU-TH 16/95\\
            IPNO/TH 95-53\\October 1995
            \end{flushleft}}
\def\eq{eq.~}

\thispagestyle{empty}
\header \vspace*{2.5cm} \centerline{\Large\bf Electromagnetic Corrections
to the decays $\eta \rightarrow 3\pi$
 \footnote{partially supported by
    Schweizerischer Nationalfonds.}}
\vspace*{2.5cm}
\centerline{\large R. Baur$^a$, J. Kambor$^b$ and D. Wyler$^a$}
\vspace*{1.0cm}
\centerline{$^a$ Theoretische Physik, Universit\"at Z\"urich, CH-8057
Z\"urich, Switzerland}
\vspace*{0.1cm}
\centerline{$^b$ Division de Physique Th\'eorique
$\footnote{Unit\'e de Recherche des Universit\'es Paris 11 et Paris 6,
associ\'ee au CNRS.}$, Institut de Physique Nucl\'eaire,}
\centerline{F-91406 Orsay Cedex, France}

\vspace*{3.5cm}
\centerline{\large\bf Abstract}
\vspace*{0.5cm}
We calculate the electromagnetic corrections to the decays
$\eta \rightarrow 3\pi$ in next to leading order in the chiral
expansion. We find that the corrections are small in accordance
with Sutherland's theorem and modify neither rate nor the
Dalitz plot distributions noticeably.

\newpage

\Section{Introduction}

In the isospin limit $m_u = m_d$ and $e = 0$,
the decay $\eta\to 3\pi$ is forbidden.
Consequently, the decay amplitude receives contributions from the
QCD isospin violating interaction
\beq
\label{QCD}
{\cal H}_{\rm QCD}(x) = \frac{(m_d-m_u)}{2}(\bar d d-\bar u u)(x),
\eeq
and from the electromagnetic interactions
\beq
\label{QED}
{\cal H}_{\rm QED}(x) = -\frac{1}{2} e^2 \int dy D^{\mu\nu}(x-y)
T(j_\mu(x) j_\nu(y)),
\eeq
where $D^{\mu\nu}(x-y)$ is the photon propagator.
Using soft pion techniques, it has been shown that the electromagnetic
contribution is much too small to account for the experimentally observed
rate \cite{SUT,BS68}.
Much work has been devoted to study the effects of the QCD contribution,
equation (\ref{QCD}).
Gasser and Leutwyler \cite{GL1}
carried out the one-loop calculation within chiral
perturbation theory \cite{WE,CH1,Leut,GL2}. Although they found large
corrections to the rate
\footnote{The Dalitz plot distribution does not get equally large corrections
from higher orders in the chiral expansion. It is therefore possibly
more sensitive to electromagnetic corrections.}
their result failed to reproduce the experimental rates \cite {PDT}.
Recently, the complete unitary corrections were evaluated \cite{KWW}.
Despite considerable uncertainties, also these corrections are not quite
sufficient to account for the observed rate, if the usually assumed value
of $(m_u-m_d)$ \cite{MASSES} is taken.  Instead, a somewhat smaller up-quark
mass
is required.

In this situation, it is of interest to reconsider the electromagnetic
corrections. Strictly speaking, they may be divided into the indirect ones,
affecting the parameters such as $F_\pi$ which enter the calculation
and the direct ones, specific for the process under consideration.
In the following we will be concerned with the latter only.
Corrections to   $F_\pi$ were calculated previously \cite{fpi} and change
its value by about 1 \%, leading to a noteworthy increase
of the eta decay rate by about 4 \%.

Within the framework of chiral perturbation
theory, electromagnetic correction can also be described by a series of
effective operators
of increasing power in momentum or masses of the mesons \cite{RES}.
Sutherland's theorem \cite{SUT} states that the first correction
of order $e^2 p^0$ vanishes where p is a typical momentum. Thus, the
electromagnetic corrections are at most of order $e^2p^2$. If the decay
amplitude is assumed to be linear in the Mandelstam variables, these
corrections are further suppressed, i.e. $p^2$ is of order $M_{\pi}^2$
rather than $M_{\eta}^2$ \cite{BS68,DDE73}.

A similar result (Dashen's theorem \cite{DT}) for the electromagnetic
mass differences states that they are equal for the pions and
for the Kaons to leading order in the chiral expansion. Recently,
it was argued that this equality could be violated \cite{DHW}
(see, however \cite{RR}),
and one
might expect then that also Sutherland's theorem receives
larger corrections than previously thought which could enhance the decay
rate of the $\eta$.
In this article we determine the ${\cal O}(e^2p^2)$ electromagnetic
corrections to the decay amplitudes $\eta\rightarrow 3\pi$. We show
that they can be safely neglected
compared to the QCD isospin violating contributions,
both for the rates and the Dalitz plot distribution.
The paper is organized as follows: in sections 2 and 3
we review the chiral lagrangian
for electromagnetic interactions up to order $p^2\alpha$. In section 4 we
calculate the electromagnetic corrections to $\eta\rightarrow 3\pi$ to one-loop
chiral perturbation theory. The numerical results are given in
section 5 and section 6 contains our conclusions.

\section{Effective low-energy Lagrangian}
To lowest order in Chiral Pertubation Theory (ChPT),
including the electromagnetic interactions, the
Lagrangian $ {\cal{L}}_{2}$ is
given by \cite{CH1,GL2,RESOS}
\begin{eqnarray}
 {\cal{L}}_{2}	 & = & -\frac{1}{4} \field \fieldu
	            - \frac{\xi}{2}(\parti \phu)^{2}
	\nonumber \\
	 & + & \frac{F^{2}_{0}}{4}\bra \dm \dmud \ket
	            +\frac{F^{2}_{0}}{4}\bra U\chid + \chi \ud \ket
	            +C\bra Q_{R}UQ_{L}\ud \ket .
\label{l2}
\end{eqnarray}
Here, $\field $ is the field strength tensor of the photon field \ph
\begin{equation}
\field = \parti a_{\nu} - \partial_{\nu} \ph .
\end{equation}
The parameter $\xi$ is
the gauge fixing parameter, which is set to $\xi = 1 $
henceforth. The pseudoscalar meson
fields are contained in the usual way in the matrix $U$,
and the field $\chi$ incorporates the coupling of the pseudoscalar mesons to
the scalar and pseudoscalar currents $s$ and $p$
\begin{equation}
	\chi = 2\bo ( s+ip) = 2\bo M +\ldots
\end{equation}
where the quark mass-matrix is
\begin{equation}
	 M=\left( \begin{array}{ccc}
	                         m_{u} &      0  &0       \\
	                         0        &m_{d} & 0       \\
	                         0        &    0    &m_{s}
	                         \end{array} \right) \quad .
\end{equation}
The covariant derivative $\dmu$ defines the coupling of the pseudoscalar
mesons to the photon field $\phu$, the external vector current $V^{\mu}$,
and the external axial current $A^{\mu}$
\begin{equation}
	\dmu = \paru U -i(V^{\mu}+Q_{R}\phu +A^{\mu}) U
	                        -iU(V^{\mu}+Q_{L}\phu -A^{\mu}).
\end{equation} \\
The two spurios fields $Q_{R}$ and $Q_{L}$ are introduced to construct a
$SU(3)_{L}\otimes SU(3)_{R}$ invariant Lagrangian. In the later
calculation they will be fixed by $Q_{R}=Q_{L}=Q$, where  $Q$ is the
charge matrix of the three light quarks
\begin{equation}
	Q = \frac{e}{3}\left( \begin{array}{ccc}
	                         2  &0  &0   \\
	                         0  &-1&0    \\
	                         0  &0  &-1
	                         \end{array} \right).
\end{equation}
The constant $F_{0}$ is up to order $O(m_{q})$ the pion decay constant.

The low energy constant $C$ determines the electromagnetic part of the
masses $M_{\pi^{\pm}},M_{K^{\pm}},$ of the charged pion and Kaon in the
chiral limit
\begin{equation}
	C\bra QUQ\ud \ket = -\frac{2e^2 C}{\fp}(\pi^{+}\pi^{-}+K^{+}K^{-})
	                                 +....
\label{cterm}
\end{equation}
Clearly, $C\bra QUQ\ud \ket$ contributes equally to the square of the
masses of $\pi^{\pm}$ and $K^{\pm}$, in agreement with Dashen's theorem
\cite{DT}.

Since the external currents $V^{\mu}$ and $A^{\mu}$ are of order $O(p)$,
where $p$ is their external momentum, the product $Q a^\mu$
must have dimension $O(p)$.
In order to formally maintain consistent chiral counting, it
is convenient to set \cite{RES}
\begin{eqnarray}
	 e \sim O(p), & \phu  \sim 1,
\end{eqnarray}
such that ${\cal L}_{2}$ is $O(p^2)$.
\\

\section{Effective Lagrangian to order $O(p^4)$}

At order $O(p^4)$ the generating functional has three different types of
contributions:
\begin{itemize}
	\item  An explicit local action of order $O(p^4)$
	\item  The one-loop graphs associated with the lowest order
	  Lagrangian ${\cal{L}}_{2}$.
	\item  The anomaly is of order  $O(p^4)$; its contributions
              will not be discussed here. For a
              discussion, the reader is referred to
              the literature \cite{CH1,GL2}.
\end{itemize}
Since we are not concerned here with the
strong intractions,\cite{GL1}, we only give the electromagnetic
terms.
The most general chiral and Lorentz invariant, P and C symmetric
Lagrangian of order $O(p^4)$ is \cite{RES}
\begin{eqnarray}
{\cal{L}}_{4} & = & K_{1} \fo \bra \dmud \dm \ket \bra Q^{2} \ket
	  +  K_{2} \fo \bra \dmud \dm \ket \bra Q U Q \ud \ket
	          \nonumber  \\
	 & + & K_{3} \fo \left( \bra \dmud Q U \ket \bra \dmd Q U \ket +
	           \bra \dmu Q \ud \ket \bra \dm Q \ud \ket \right)
	            \nonumber   \\
	 & + & K_{4} \fo \bra \dmud Q U \ket \bra \dm Q \ud \ket
	           + K_{5} \fo  \bra \left\{\dmud ,\dm  \right\} Q^2 \ket
	          \nonumber  \\
	 & + & K_{6} \fo \bra \dmud \dm  Q \ud Q U +\dmu \dmd Q U Q \ud
	           \ket
	          \nonumber   \\
	 & + & K_{7} \fo \bra \chid U + \ud \chi \ket \bra Q^2 \ket
	          + K_{8}\fo \bra \chid U + \ud \chi \ket \bra Q U Q \ud \ket
	           \nonumber  \\
	 & + & K_{9}\fo \bra \left( \chid U + \ud \chi \right) Q^2
	                            +\left( \chi \ud + U \chid \right) Q^2\ket
	           \nonumber   \\
	 & + & K_{10} \fo \bra \left( \chid U + \ud \chi \right) Q \ud Q U +
	           \left( \chi \ud + U \chid \right) Q U Q \ud \ket
	           \nonumber   \\
	 & + & K_{11} \fo \bra \left( \chid U - \ud \chi \right) Q \ud Q U +
	           \left( \chi \ud - U \chid \right) Q U Q \ud \ket
	           \nonumber  \\
	 & + &  K_{12} \fo \bra \dmud \left[ \qm{R},Q \right] U
	           + \dmu \left[\qm{L}, Q \right] \ud \ket
	           \nonumber  \\
	 & + & K_{13}\fo \bra \qmu{R} U \qm{L} \ud \ket +
	            K_{14}\fo \bra \qmu{R} \qm{R} + \qmu{L}\qm{L} \ket
	          \nonumber  \\
	 & + & K_{15} F^{4}_{0} \bra Q U Q \ud \ket^2 +
	          K_{16} F^{4}_{0} \bra Q U Q \ud \ket \bra Q^2 \ket +
	          K_{17} F^{4}_{0} \bra Q^2 \ket^{2},
\label{lag4}
\end{eqnarray}
where $F^{R}_{\mu \nu}$ and  $F^{L}_{\mu \nu}$ are the field
strength tensors of $ F^{R}_{\mu}$ and of $ F^{L}_{\mu}$ respectively
\begin{eqnarray}
F^{R}_{\mu}& =& V_{\mu}+Q^{R}\ph + A_{\mu} \nonumber \\
F^{L}_{\mu}& =& V_{\mu}+Q^{L}\ph - A_{\mu} \nonumber \\
F^{I}_{\mu \nu}& =& \parti F^{I}_{\nu}-\partial_{\nu} F^{I}_{\mu} -i\left[
	F^{I}_{\mu} , F^{I}_{\nu} \right] ,  \quad I=R,L
\end{eqnarray}
and the covariant derivative \qm{I} is defined as
\begin{equation}
\qm{I} = \parti Q^{I} +i \left[Q^{I},F^{I}_{\mu}\right], \quad  I=R,L.
\end{equation}
Under a chiral transformation the covariant derivative \qm{I} transforms
like the corresponding $Q^{I}$
\begin{eqnarray}
 \qm{I} & \rightarrow & V_{I}\qm{I}V^{\dagger}_{I}  \quad I=R,L
 \nonumber \\
	V_{R,L} & \epsilon & SU(3)_{R,L} \quad.
\end{eqnarray}
We have set $Q^{R}=Q^{L}=Q $. Only in the covariant derivative $\qm{I}$
have we kept the index $ I=R,L $ in order to make explicit the proper chiral
transformation. The coupling constants
$K_{1} \ldots K_{17}$
are not determined by chiral symmetry.
While in principle
calculable from the fundamental lagrangian of QCD and QED, we
consider them as phenomenological
constants.
$K_{1},K_{7},K_{16}$ are electromagnetic corrections to $\fo$, $\bo$ and
$C$
respectively.

In the next step, the loops generated by the Lagrangian  ${\cal L}_{2}$
are calculated. They lead to terms proportional to $p^4$. Their divergencies
are cancelled by infinite counterterms in the constants $K_{i}$. The
procedure is standard and will not displayed here in detail \cite{RES}.
Using the usual dimensional regularization scheme with
scale $\mu$, one finds that the
ultraviolet divergencies can be absorbed in
 the coupling constants $K_i$ with the following
 renormalization ot the low-energy coupling constants\cite{RES}:
 \begin{eqnarray}
  	K_i & = & K^{r}_{i}(\mu)+\Sigma_i \lambda  \nonumber \\
\lambda & = & \frac{\mu^{d-4}}{16\pi^2}\left\{
 	\frac{1}{d-4}-\frac{1}{2}\left[\ln (4\pi) +\Gamma^{'}(1)+1\right]
 	\right\}
\label{counter}
 \end{eqnarray}
and
 \begin{equation}
\begin{array}{lll}
  	\Sigma_{1}=\frac{3}{4} & \Sigma_{2}=Z  &
 	\Sigma_{3}=-\frac{3}{4}
 	\nonumber \\ \\
 	\Sigma_{4}=2Z  & \Sigma_{5}=-\frac{9}{4}  &
 	\Sigma_{6}=\frac{3}{2}Z
 	\nonumber \\ \\
 	\Sigma_{7}=0  & \Sigma_{8}=Z  &
 	\Sigma_{9}=-\frac{1}{4}
 	\nonumber \\ \\
 	\Sigma_{10}=\frac{3}{2}Z+\frac{1}{4}  &
 	\Sigma_{11}=\frac{1}{8}  &
 	\Sigma_{12}=\frac{1}{8}
            \nonumber \\ \\
 	\Sigma_{13}=0  & \Sigma_{14}=0  &
 	\Sigma_{15}=20Z^2+3Z+ \frac{3}{2}  \nonumber \\  \\
 	\Sigma_{16}=-4Z^2 - \frac{3}{2}Z -3  &
 	\Sigma_{17}=2Z^2 - \frac{3}{2}Z +\frac{3}{2} & \\ \\
	Z=\frac{C}{F^{4}_{0}} & &
\label{sigma1}
\end{array}
\end{equation}
\\
Here, the $K^{r}_{i}(\mu)$ are the renormalized couplings at the
scale $\mu$.

The scaling behaviour of  $K^{r}_{i}(\mu)$ is determined by the
requirement
\begin{equation}
\mu\frac{dK_{i}}{d\mu} = 0 \nonumber
\end{equation}
which implies for  $K^{r}_{i}(\mu)$  in the limit
$d=4$
\begin{equation}
K^{r}_{i}(\mu) = K^{r}_{i}(\mu_{0}) - \frac{\Sigma_{i}}{16\pi^{2}} \ln(
                                                     \frac{\mu}{\mu_{0}}).
                                                     \label{group}
\end{equation}

\section{Electromagnetic contributions to $ \eta \to 3\pi $}

The electromagnetic contributions to the decay $\eta\rightarrow 3\pi$ have
been considered long ago \cite{BS68,DDE73}. The transition amplitude is
suppressed due to a soft-pion theorem \cite{SUT}. At next-to-leading order
in a low energy expansion, there is a further suppression factor
$m_\pi^2/m_K^2$, if linear dependence on the energy of the odd pion is
assumed. This led to the conclusion that the electromagnetic interactions
alone fail completely to explain the observed rate.

Here the problem is reinvestigated in the framework of effective chiral
lagrangians, as reviewed in section 2. We calculate the corrections of order
$e^2 p^2$, where $p^2$ is a generic low energy momentum. We work in the
isospin limit, as corrections of order $e^2 (m_d-m_u)$ are very small.

As usual, we define the decay amplitude $A$ by
\begin{eqnarray}
< \! \pi^{0} \pi^{+} \pi^{-}\mbox{out}\mid \eta \! >
      & = &i\ (2 \pi^{4} )\ \delta^{4}(P_{f}-P_{i})A(s,t,u),
\end{eqnarray}
with the Mandelstam variables
\begin{equation}
s=(p_{\eta}-p_{ \pi^{0}})^{2} \quad t=(p_{\eta}-p_{ \pi^{+}})^{2}
\quad u=(p_{\eta}-p_{ \pi^{-}})^{2}. \nonumber
\end{equation}

Due to G-parity, the three pions emerge in a $I=1$ state. Hence, the decay
amplitude into three neutral pions, $\bar A(s,t,u)$, is also determined
by $A(s,t,u)$
\begin{eqnarray}
< \! \pi^{0} \pi^{0} \pi^{0}\mbox{out}\mid \eta \! >
      & = &i\ (2 \pi^{4} )\ \delta^{4}(P_{f}-P_{i})\ \bar A(s,t,u)
\nonumber \\
\bar A(s,t,u)  & = & A(s,t,u) +A(t,u,s) +A(u,s,t) .
\label{eqone}
\end{eqnarray}
Charge conjugation invariance requires $A(s,t,u)$ to be symmetric with respect
to interchanges of $t$ and $u$.

The electromagnetic contributions to $A(s,t,u)$ may be written as
\begin{equation}
A_{QED}(s,t,u) = e^2 f(\Lambda,p^2,m_{u},m_{d},,m_{s},\ldots) \\
\end{equation}
where $\Lambda$ is the scale of the theory and $p^2$ stands for any of
the variables s,t,u. Keeping ratios $p^2$:$m_u$:$m_d$:$m_s$ fixed, $f$ can be
expanded according to
\begin{equation}
 f = f_{0}+f_{1}+\ldots \quad .
\end{equation}
$f_0$ is of order $p^0$, $f_1$ of order $p^2$ etc. The lowest order term
$f_0$ vanishes, as was shown by Sutherland \cite{SUT}. As noted,
 $f_1$
is also suppressed, if the
energy dependence of the amplitude is linear.
This follows from the soft-pion theorem \cite{SUT}
\begin{equation}
\lim_{p_0\rightarrow 0} A_{\rm QED}^{+-0}=
\lim_{p_0\rightarrow 0} A_{\rm QED}^{000}={\cal O}(e^2 \hat m).
\end{equation}
Hence, for a linear amplitude
\begin{equation}
A_{\rm QED}^{+-0}=a+b\cdot s,
\label{linapp}
\end{equation}
it follows that at the soft-pion point $s=M_\eta^2+3 M_\pi^2$, and $t=u=0$
\begin{eqnarray}
A_{\rm QED}^{+-0}(p_0=0)&=&a+b (M_\eta^2+3 M_\pi^2) \\
A_{\rm QED}^{000}(p_0=0)&=&3a+b (M_\eta^2+3 M_\pi^2),
\end{eqnarray}
and therefore
\begin{equation}
a=0, \quad b={\cal O}({\hat m\over m_s}) .
\end{equation}

In chiral perturbation theory the term $f_1$ can be calculated in a systematic
manner. The assumption (\ref{linapp}) is seen to be violated by non-local terms
arising from one-loop graphs. However, only Kaons  appear in the loop,
therefore the resulting amplitude is smooth throughout the physical region.
Equation (\ref{linapp}) is then still a good approximation and the
${\hat m \over m_s}$
suppression is effectively at work. This is the main reason why the
next-to-leading order term $f_1$ is small.

The relevant diagrams are shown in figure 1. The graphs a) and b) are the
non-local unitary and the tadpole graphs respectively, and c) contains the
direct and pole graphs.
\clearpage
\begin{figure}[!t]
 \begin{center}
 \begin{tabular}{c}
   \epsfxsize=15.0cm
   \leavevmode
   \epsffile[0 250 550 850]{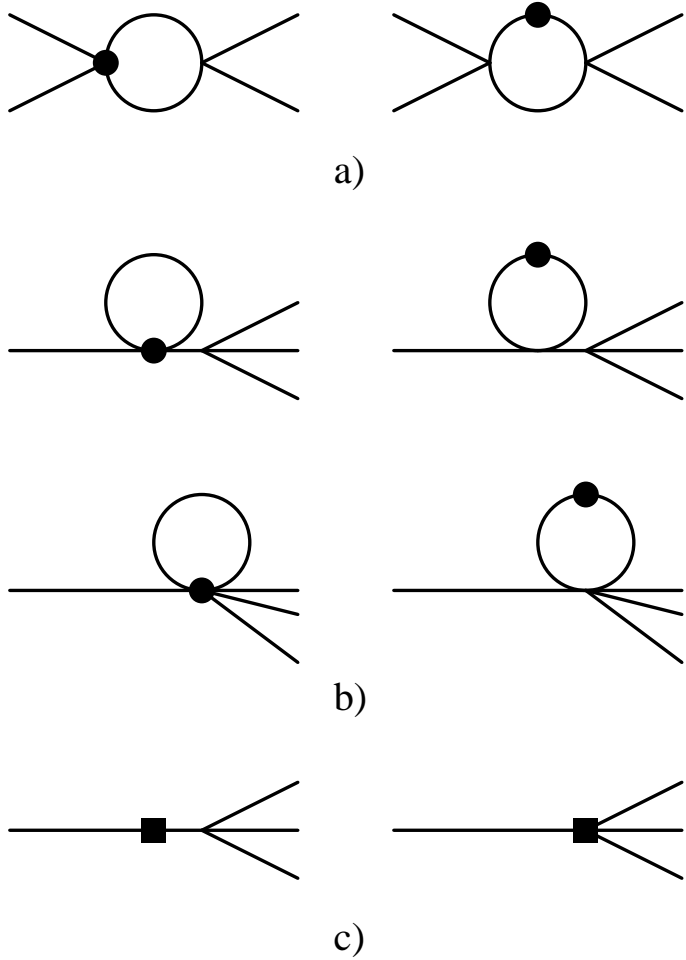}
 \end{tabular}
 \caption[]{\label{plot}Graphs contributing to the decay  $\eta \to 3\pi$.
 The filled circles denote vertices of order $O(e^2)$ and the filled boxes
those of order $O(e^2 p^2)$.  }
 \end{center}
\end{figure}

The electromagnetic vertex in these diagrams is given by the leading
order term in eq. (\ref{l2}) and by eq. (\ref{lag4}). There are no graphs
with pions in the
loop as they are in addition suppressed by the
$\pi^0 -\eta$ mixing angle $\ep$, which is of the order
$$
O(\ep) \approx 10^{-2} =O(\frac{m_{u}-m_{d}}{m_{s}}).
$$
Also, the mass difference $(M^{2}_{K^{\pm}}-M^{2}_{K^{0}})_{QCD}$ contributes
only to the order $ O(\frac{m_{u}-m_{d}}{m_{s}}) $ and can be neglected.

We then obtain for the unitary corrections
\begin{eqnarray}
U(s,t,u) = \frac{C}{12 \sqrt{3} F^{6}_{0}} \lbrack
           3s(3s-4M^{2}_{K})\frac{1}{i}C_{kk}(s) +
6(3s-4M^{2}_{K})J_{kk}(s)
 \nonumber  \\
            -3(3t-4M^{2}_{K})J_{kk}(t)-3(3u-4M^{2}_{K})J_{kk}(u)
\nonumber \\
	    -(5s-4M^{2}_{K})J_{kk}(0) \rbrack ,
\end{eqnarray}
 for the tadpoles
\begin{eqnarray}
T(s) &=& -\frac{C}{ \sqrt{3} F^{6}_{0}} \lbrack
       (s-s_{0}) (\frac{9}{4}
+\frac{2M^{2}_{\pi}}{M^{2}_{\eta}-M^{2}_{\pi}})
         \nonumber \\
   & &   + \frac{2M^{2}_{\pi}}{3}-\frac{5s}{12}
+\frac{M^{2}_{K}}{3}\rbrack  J_{kk}(0) ,
\end{eqnarray}
and finally for the  sum of the pole and direct graphs
\begin{eqnarray}
D(s) &= &-\frac{4 M^{2}_{\pi}}{9 \sqrt{3} F^{2}_{0}} \lbrack
         1+\frac{3(s-s_{0})}{M^{2}_{\eta}-M^{2}_{\pi}}\rbrack \times
            \nonumber \\
       & &    \{ -3K_{3}+\frac{3}{2}K_{4}+  K_{5}+K_{6}
            - K_{9}- K_{10} \} .
\end{eqnarray}
We have introduced the following functions $C_{kk}$ and $J_{kk}$ :
\begin{eqnarray}
\frac{1}{i}C_{kk}(s)
&=&-\frac{1}{16 \pi^2} \frac{1}{M^{2}_{K}} \frac{2}{\sqrt \Delta}
	 \arctan \frac{ \bar s }{\sqrt \Delta}   \nonumber \\
J_{kk}(s) &=& \frac{1}{16 \pi^2} \lbrack 2- 2\frac{\sqrt \Delta}{ \bar s }
           \arctan \frac{ \bar s }{\sqrt \Delta} \rbrack + J_{kk}(0)
\nonumber \\
J_{kk}(0) &=& -2\lambda -2k + O(d-4)
\end{eqnarray}
with
$$
 \bar s = \frac{s}{M^{2}_{K}} \qquad \Delta = \bar s (4- \bar s) \nonumber
$$
\begin{equation}
3s_{0} = s+t+u =m^{2}_{\eta}+3m^{2}_{\pi} ,
\end{equation}
and
\begin{eqnarray}
\lambda & = & \frac{\mu^{d-4}}{16 \pi^2} \{ \frac{1}{d-4}-\frac{1}{2}
               (\ln(4\pi)+\Gamma'(1)+1) \}  \nonumber \\
 k   & = & \frac{1}{32 \pi^2} \{ \ln\frac{M^{2}_{K}}{\mu^2} +1 \}.
\end{eqnarray}
With  equation (\ref{counter})
and the following definition:
\begin{equation}
\bar J_{kk}(s)=J_{kk}(s)-J_{kk}(0)
\end{equation}
we obtain the finite result
\begin{eqnarray}
A_{QED}(s,t,u)&=&e^2 (U(s,t,u)+T(s)+D(s) ) \nonumber \\
&=&  \frac{Ce^2}{ \sqrt{3} F^{6}_{0}} \lbrack
\frac{1}{4}s(3s-4M^{2}_{K})\frac{1}{i}C_{kk}(s)
+\frac{1}{2}(3s-4M^{2}_{K})\bar J_{kk}(s)  \nonumber  \\
& &  -\frac{1}{4}(3t-4M^{2}_{K})\bar J_{kk}(t)
    -\frac{1}{4}(3u-4M^{2}_{K})\bar J_{kk}(u) \rbrack \nonumber \\
  & &+\frac{4Ce^2M^{2}_{\pi}}{3 \sqrt{3} F^{6}_{0}}
\{ k - \frac{F^{4}_{0}}{3C} \bar K^{r}(\mu )\}
  \lbrack
1+\frac{3(s-s_{0})}{M^{2}_{\eta}-M^{2}_{\pi}}\rbrack
\label{finite}
\end{eqnarray}
with
\begin{eqnarray}
\bar K^{r}(\mu)&=& -3K^{r}_{3}(\mu)+\frac{3}{2}K^{r}_{4}(\mu)
                             +  K^{r}_{5}(\mu) \nonumber \\
                      &+&  K^{r}_{6}(\mu) - K^{r}_{9}(\mu)-
K^{r}_{10}(\mu).
\end{eqnarray}
Some remarks concerning the structure of this result are in order:
\begin{enumerate}
\item
The scale dependence of $k$ is cancelled by the running of $\bar K^{r}(\mu)$.
Using equations (\ref{sigma1}) and (\ref{group}) one gets
\begin{equation}
\bar K^{r}(\mu) =\bar K^{r}(\mu_{0}) -\frac{3C}{F^4_0 32\pi^2}
\ln(\frac{\mu^2}{\mu^2_{0}}).
\end{equation}

\item
As expected, the amplitude vanishes in the soft pion point
\begin{equation}
	s \to  M^{2}_{\eta}, \qquad t,u  \to  0
\end{equation}
if the lowest order relation $3 m_\eta^2= 4 m_K^2$ is used. This is
in accord with Sutherland's theorem which predicts a vanishing amplitude
if the pion mass is sent to zero.

\item
The final result (\ref{finite}) consists of a nonlocal piece (first two
lines) and a polynomial to first order in s (last line). The polynomial
part is in proportion to the current algebra amplitude of strong interactions.
It exhibits the
$M_\pi^2$ suppression as implied by Sutherlands theorem. Since the counterterm
$\bar K^r$ enters only this part of the amplitude, it's contribution is
suppressed too. The nonlocal piece circumvents this suppression as it is
clearly not linear in s. However, since only Kaon loops contribute, this
part is kinematically suppressed.

\item
The rate is to a large extent fixed by the amplitude at the center of the
Dalitz plot. We therefore expand (\ref{finite}) according to
\begin{equation}
A_{\rm QED}(s,t,u)=a_0+a_1 (s-s_0)+a_2 (s-s_0)^2+a_3 (t-u)^2 .
\end{equation}
Explicitly, the constant term is given by
\begin{equation}
a_0={C e^2\over 3\sqrt{3} F_0^6} \left\{
{3\over 4} s_0 (3 s_0-4 M_K^2) {1\over i} C_{kk}(s_0)
+4 M_\pi^2 \left( k-{F_0^4 \over 3 C} \bar K^r(\mu) \right) \right\}.
\end{equation}
Note that at $s=s_0$ the remaining nonlocal piece in proportion to
$C_{kk}$ is numerically less suppressed than for instance $\bar J_{kk}$.

\item
The linear slope $a_1$ is given in terms of the same counterterm $\bar K^r$
as is the amplitude at the center of the physical region. Thus, if isospin
breaking in the quark masses is switched off, the
electromagnetic rate would be fixed in terms of the linear slope.
The quadratic coefficients $a_2$, $a_3$ get contributions only from the
nonlocal part of $A_{\rm QED}$. They are given in terms of the known
coupling constant $C$.
\end{enumerate}

\section{Numerical Results}
As mentioned, the one-loop result of Gasser and Leutwyler yields
a rate of $167$ eV, much below the experimental value. Although
the unitary corrections are available \cite{KWW}, we will use the
one-loop result as reference. This is inessential, since the
electromagnetic corrections are very small.

We set $F_{0}$ equal to the pion decay constant $F_{\pi}$, whose
experimental value is  $F_{\pi} = 92.4$ MeV  if
electromagnetic corrections are included \cite{fpi}.
The low energy constant $C$
can be determined by
using the underlying vector and axial-vector resonances. Using the
Weinberg sum
rules \cite{WE2}, one obtains \cite{RESOS}
\begin{equation}
C= \frac{3}{32\pi^2}M^{2}_{\rho}F^{2}_{\rho}\ln(\frac{F^{2}_{\rho}}{
      F^{2}_{\rho}-\fp}),
\end{equation}
where $M_{\rho} $ is the mass of the $\rho$-meson, $M_{\rho}=770$ MeV and
$F_{\rho}$ is its decay constant, $F_{\rho}=154$ MeV. Numerically,
\begin{equation}
{C \over F_\pi^4}= 0.84.
\end{equation}
The values of the coupling constants $K^{r}_{i}$ are so
far unknown. The `naive' chiral counting law \cite{MG84} gives
\begin{equation}
\label{naive}
	\mid K^{r}_{i} \mid = \frac{1}{16\pi^2}\approx 6.3 \times 10^{-3},
\end{equation}
while an estimate  for $K^{r}_{8}$ yields \cite{RES}
\begin{equation}
	K^{r}_{8}(\mu=0.5GeV) = (-1.0\pm 1.7)\times 10^{-3},
\end{equation}
consistent with the naive rule (\ref{naive}).

We conservatively use eq. (\ref{naive}) and vary the
absolute value of $\bar K^r$ between the bounds
\begin{equation}
\label{vari}
\mid \bar K^r \mid\leq \frac{8.5}{16\pi^2} .
\end{equation}

To get a feeling of the numerical size of the elecromagnetic corretions, we
expand the amplitude (\ref{finite}) around the center of the Dalitz plot
and compare it to the ChPT one-loop result for the strong interactions.
Writing the Taylor expansion of the sum of these amplitudes as
\begin{equation}
A_{\rm QCD}(s,t,u)+A_{\rm QED}(s,t,u)=a_0+a_1 (s-s_0)+a_2 (s-s_0)^2+a_3 (t-u)^2
\end{equation}
we find
\begin{eqnarray}
a_{0}&=&a^{QCD}_{0}+a^{n.l}_{0}+a^{l}_{0} \nonumber \\
&= & -(0.17+0.02i) +0.14 \times 10^{-2} \pm 0.30 \times 10^{-2} ,
\end{eqnarray}
where $a^{QCD}_{0}$ denotes the strong one-loop value, $a^{n.l}_{0}$
the contribution from the non-local terms in eq.(\ref{finite}), and
$a^{l}_{0}$ that of the local terms which depend on the low energy coupling
constants $K_{i}$. The variation of
$a^{l}_{0}$ are due to the changing value of $\bar K^r$.
Similarly, we obtain
\begin{eqnarray}
a_{1}&=&a^{QCD}_{1}+a^{n.l}_{1}+a^{l}_{1} \nonumber \\
     &=&-(2.25+0.75i)+0.62 \times 10^{-3} \pm 0.30 \times 10^{-1}
 \quad [{\rm GeV}^{-2}] . \\
a_{2}&=&a^{QCD}_{2}+a^{n.l}_{2} \nonumber \\
     &=&(1.33-0.91i)-0.25 \times 10^{-1}
\quad [{\rm GeV}^{-4}] . \\
a_{3}&=&a^{QCD}_{3}+a^{n.l}_{3} \nonumber \\
     &=&-(0.93-0.41i)-0.85 \times 10^{-2}
 \quad [{\rm GeV}^{-4}] .
\end{eqnarray}
Here, we used everywhere the `new' value of $F_\pi$ \cite{fpi}; the
other relevant factors for the QCD amplitude are taken from ref \cite{GL1}.
 From this we see that for the extreme values of  $\bar K^r$ the local
contribution dominates over the non-local contribution. Both are, however,
small compared to the ChPT one-loop result. We therefore expect no significant
changes from QED to both, the decay rates as well as the Dalitz plot
distributions
for the decays $\eta \to \pi^{0} \pi^{+}\pi^{-}$ and $\eta \to 3 \pi^0$.

The decay rate $\Gamma_{\eta \to \pi^{0} \pi^{+}\pi^{-}}$ is
obtained by evaluating the phase space integral
\footnote{Note that the phase space is very small and thus very sensitive
to the
mass difference $M_{\pi^{\pm}} - M_{\pi^0}$}
	\begin{eqnarray}
		d\Gamma_{\eta \to \pi^{0} \pi^{+}\pi^{-} }	 & = &
\frac{(2\pi)^4}{M_{\eta}} \delta (P_{i}- P_{f})
\mid  A_{QCD}(s,t,u)+A_{QED}(s,t,u)\mid^2  d\mu ,
		\nonumber  \\
d\mu &=&\frac{d^3 p_{\pi^{0}}}{(2\pi)^4 2p^{0}_{\pi^{0}}}
        \frac{d^3 p_{\pi^{+}}}{(2\pi)^4 2p^{0}_{\pi^{+}}}
        \frac{d^3 p_{\pi^{-}}}{(2\pi)^4 2p^{0}_{\pi^{-}}} .
	\end{eqnarray}
We get for the integrated rate
\begin{equation}
\Gamma_{\eta \to  \pi^{0} \pi^{+}\pi^{-} } = 165 \pm 5 {\rm eV}
\end{equation}
where the uncertainty is due to our lack of knowledge on $\bar K^r$, and
for the ratio $r$
\begin{equation}
	1.42\leq	r=\frac{\Gamma_{\eta \to  3\pi^{0}}}
		{\Gamma_{\eta \to  \pi^{0} \pi^{+}\pi^{-} }}
		\leq 1.43 \quad .
\end{equation}
In both cases elecromagnetism gives, as expected, small corrections to the
results
obtained by \cite{GL1}, where the purely strong effects have been considered
\begin{eqnarray}
\Gamma^{QCD}_{\eta \to  \pi^{0} \pi^{+}\pi^{-} }& = &167 {\rm eV} \\
r^{QCD}      & = & 1.43 \quad .
\end{eqnarray}
Even though QED can give rise to a corretion of up to 4 per cent to the
decay rate  $\Gamma_{\eta \to \pi^{0} \pi^{+}\pi^{-}}$,
the Dalitz plot distribution remains unchanged. This is due to the fact that
the non-local contributions are small and that the functional dependence
of the local term in eq.(\ref{finite}) on the Mandelstamm variable $s$  and
that of the current algebra amplitude are the same.
Since the Dalitz plot distribution  is a measure of the $s$-dependence
of the amplitude $A(s,t,u)$, the  $ \bar K^r (\mu_{0})$ contribution only leads
to a different normalization of the distribution, but there
is no significant change of the slopes.

\Section{Conclusions}

In this paper we have calculated the direct electromagnetic corrections
to the decay  $\eta \rightarrow 3\pi$. Using the general framework
of chiral perturbation theory, we determine the first nontrivial term of
order $e^2p^2$. This includes mesonic loops and a combination of
counterterms introduced earlier \cite{RES}. Unlike for meson mass
differences, the corrections are tiny, and amount to at
most two percent of the ChPT one-loop amplitude at the center of the
decay region.
\footnote{There are other, indirect, electromagnetic corrections
which influence the fundamental parameters, such as $F_\pi$. They
are known \cite{fpi} and change the normalization of the QCD amplitude
noticeably, resulting in a 8 \% increase of the rate.}
The Dalitz plot parameters are also rigid with respect to electromagnetic
corrections; from the results in section 5 we see that both
linear and quadratic slope parameters are practically unaffected by
the electromagnetic corrections.
This  is clear from the form of the amplitude:
First, the nonlocal part which circumvents the
Bell-Sutherland suppression arises from Kaon loops, which are kinematically
stongly suppressed. The larger (by a factor of 13) pion loops multiply the
small
quantity $(m_d-m_u)$ and can be neglected. Furthermore, the polynomial part
is suppressed by the small factor $M_\pi^2/M_K^2$. This includes in particular
the contributions from the coupling constants $K_i$.
Thus, the dependence on these little known constants is weak; varying
them between the rather large values $\pm \frac{8.5}{16 \pi^2}$
results only in a variation of less than two percent relative to the
one-loop ChPT amplitude of the strong interactions.

Thus, our result confirms the usual picture, namely that the
large decay rate requires rather large isospin violations in the strong
interactions. Moreover, any precise measurement of the
$\eta \rightarrow 3\pi$ Dalitz plot yields important information on the
strong interactions without contamination from the electromagnetic ones.

\vspace{0.5cm}

{\bf Acknowledgements}.

We thank R. Urech  for discussions and G.-J. Van Oldenborgh for the
help in the numerical part. J.K. would like to acknowledge support from
the Department of Physics and Astronomy, University of Massachusetts,
where the work described herein was started.

\end{document}